# Photo-induced Superconducting State with Long-lived Disproportionate Band Filling in FeSe

Takeshi Suzuki[1,*], Takashi Someya[1], Takahiro Hashimoto[1], Shoya Michimae[1], Mari Watanabe[1], Masami Fujisawa[1], Teruto Kanai[1], Nobuhisa Ishii[1], Jiro Itatani[1], Shigeru Kasahara[2], Yuji Matsuda[2], Takasada Shibauchi[3], Kozo Okazaki[1,*], and Shik Shin[1,*]

[1]*Institute for Solid State Physics, The University of Tokyo, Kashiwa, Chiba 277-8581, Japan*
[2]*Department of Physics, Kyoto University, Kyoto 606-8502, Japan*
[3]*Department of Advanced Materials Science, University of Tokyo, Kashiwa, Chiba 277-8561, Japan*

Photo-excitation is a very powerful way to instantaneously drive a material into a novel quantum state without any fabrication, and variable ultrafast techniques have been developed to observe how electron-, lattice-, and spin-degrees of freedom change. One of the most spectacular phenomena is photo-induced superconductivity, and it has been suggested in cuprates that the transition temperature *Tc* can be enhanced from original *Tc* with significant lattice modulations. Here we show another photo-induced high-Tc superconducting state in the iron-based superconductor FeSe with semi-metallic hole and electron bands. The transient electronic state in the entire Brillouin zone is directly observed by the time- and angle-resolved photoemission spectroscopy using extreme ultraviolet pulses obtained from high harmonic generation. Our results of dynamical behaviors on timescales from 50 fs to 800 ps consistently support the favorable superconducting state after photo-excitation well above *Tc*. This finding demonstrates that multiband iron-based superconductors emerge as an alternative candidate for photo-induced superconductors.

Among Fe-based superconductors, FeSe has attracted enormous interest from the aspects of absence of antiferromagnetism in contrast to other Fe-based superconductors, and potential superiority in exhibiting higher critical temperature ($T_c$) of superconductivity under various external applications. Although $T_c$ is only around 10 K in the ambient pressure [1], significant increase of $T_c$ has been achieved to ~ 40 K by physical pressure [2-4]. Regarding fabrications, intercalation of a spacer layer can increase $T_c$ to ~ 40 K [5,6], and $T_c$ of single-layer FeSe has been reported to exhibit about 60 K [7]. The key ingredient for achieving higher $T_c$ lies in designing the band structure. For example, dramatic increase of $T_c$ for single-layer FeSe is accompanied by the disappearance of the hole Fermi surface (FS) and increase of the electron FS [7]. In this context, photo-excitation has substantial advantages over other methods because it can instantaneously manipulate a material of interest *in situ* without any fabrication [8-10]. One of the striking phenomena is the photo-induced superconductivity reported in high-$T_c$ cuprate superconductors, where the key mechanism is played by the lattice modulation [11]. Furthermore, it should be emphasized that near-infrared pulse has also been employed for excitation, which initially excite the electronic system while the mid-infrared can resonantly excite the lattice system [12,13].

Recently, photo-excited phenomena for Fe-based superconductors have been intensively studied. In photo-excited BaFe$_2$As$_2$, the temporally periodic formation of spin density wave was observed by terahertz (THz) spectroscopy [14]. The demonstration of chemical potential control [15] and the possibility of photo-induced superconductivity [16] were reported by time- and angle-resolved photoemission spectroscopy

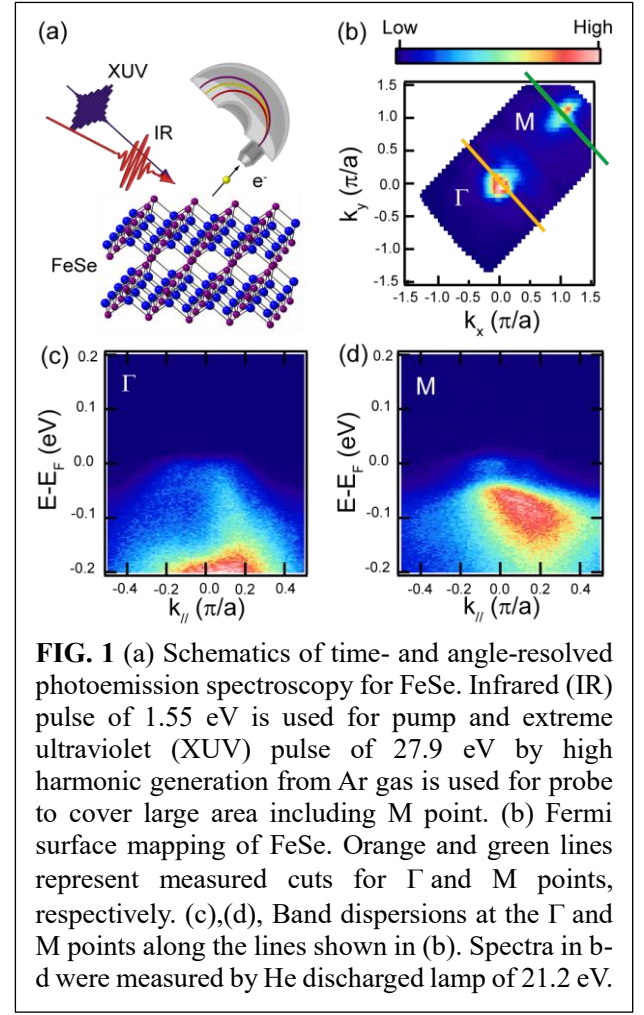

**FIG. 1** (a) Schematics of time- and angle-resolved photoemission spectroscopy for FeSe. Infrared (IR) pulse of 1.55 eV is used for pump and extreme ultraviolet (XUV) pulse of 27.9 eV by high harmonic generation from Ar gas is used for probe to cover large area including M point. (b) Fermi surface mapping of FeSe. Orange and green lines represent measured cuts for Γ and M points, respectively. (c),(d), Band dispersions at the Γ and M points along the lines shown in (b). Spectra in b-d were measured by He discharged lamp of 21.2 eV.

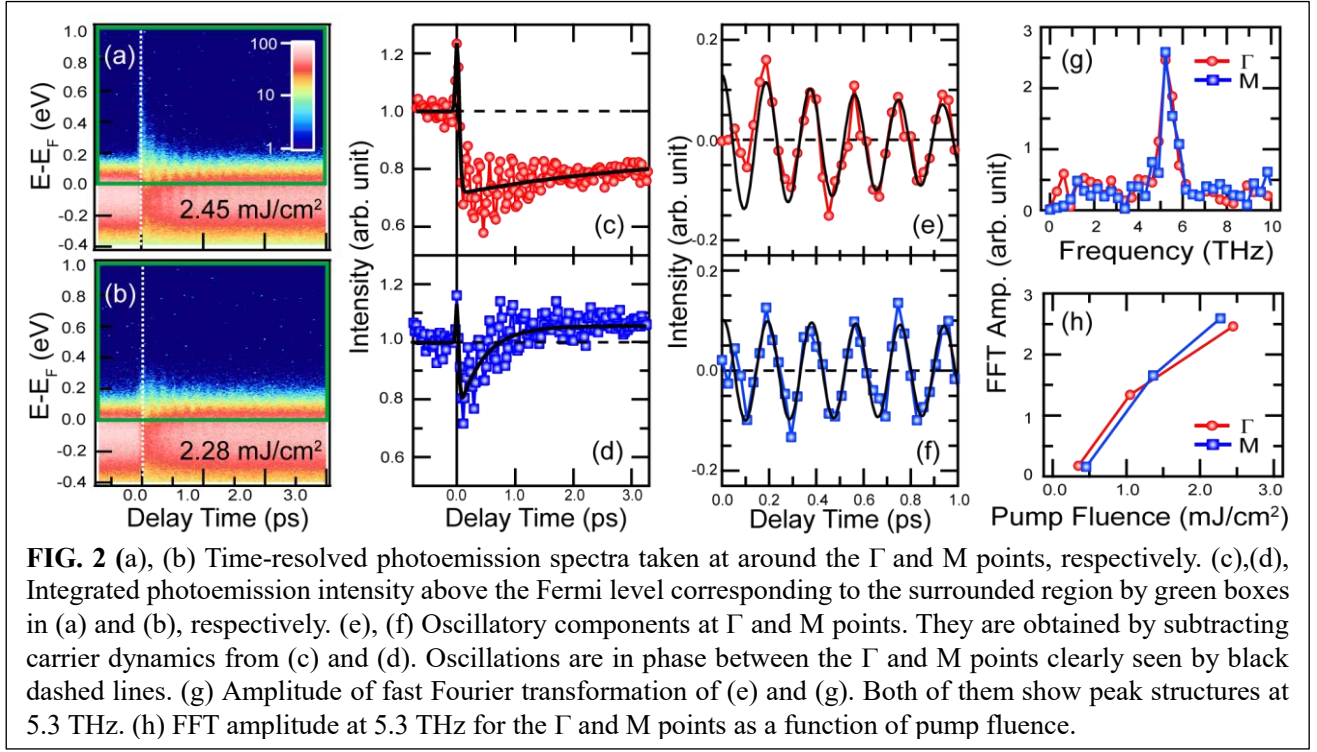

**FIG. 2** (a), (b) Time-resolved photoemission spectra taken at around the Γ and M points, respectively. (c),(d), Integrated photoemission intensity above the Fermi level corresponding to the surrounded region by green boxes in (a) and (b), respectively. (e), (f) Oscillatory components at Γ and M points. They are obtained by subtracting carrier dynamics from (c) and (d). Oscillations are in phase between the Γ and M points clearly seen by black dashed lines. (g) Amplitude of fast Fourier transformation of (e) and (g). Both of them show peak structures at 5.3 THz. (h) FFT amplitude at 5.3 THz for the Γ and M points as a function of pump fluence.

(TARPES). The key mechanism is the connection of the electron and spin properties to the lattice motion and each report is discussed along the clear appearance of coherent phonons [17, 18]. In photo-excited FeSe, coherent phonons were also reported by TARPES [19], and the connection between dynamical band shift and lattice dynamics was directly measured by combining TARPES and time-resolved x-ray diffractions [20]. However, an important but lacking aspect in photo-excited FeSe is dynamical properties for the electron pockets located around the Brillouin-zone (M point). By directly observing how the hole and electron pockets change by photo-excitation, we can also gain insight into the response of the lattice degree of freedom in the photo-excited state. Because it is a relatively easy setup, photon energy of 6 eV is typically employed for TARPES measurements. However, due to the lack of photon energy, TARPES using 6 eV cannot access the M point. This is overcome by using high harmonic generation (HHG) to obtain higher photon energy and capture a larger region of Brillouin zone [21].

Here, we investigate the non-equilibrium electronic structure of FeSe by performing TARPES using an extreme ultraviolet (XUV) laser, which is schematically shown in Fig. 1(a). Oscillations as a result of $A_{1g}$ coherent phonon excitation are clearly observed for both the hole and electron FSs, and they are found to be in phase with each other, whose behavior is different from $BaFe_2As_2$ [16]. From long-delay time measurements, we elucidate that the disproportionate band filling between the hole and electron bands, which mimics the electronic structure of a single-layer FeSe, appears and persists for longer than 800 ps. Furthermore, the additional leading edge midpoint found in both the hole and electron FSs are considered to be ascribed to the superconducting gap. By comparing with band-structure calculations, we find that the distance between neighboring Se- and Fe-layers increases in a photo-excited metastable state. This result suggests that fairly long-lived nonequilibrium electronic and lattice structures are available by photo-excitation in FeSe owing to its indirect semimetallic band structure. We propose that a long-lived photo-induced superconducting state could be expected for FeSe.

**Displacive excitation of coherent phonons**

In order to characterize the cleaved surfaces, we first performed conventional static ARPES using He discharge lamp (21.2 eV). Figure 1(b) shows the result of FS mapping, and two FSs are clearly seen around the Γ (0, 0) and M (π/a, π/a) points. Figures 1(c) and 1(d) show the ARPES intensity as a function of energy and momentum taken at around the Γ and M points, respectively. The horizontal momentum axis corresponds to the solid lines in Fig. 1(b) while the vertical energy axis is represented with respect to the Fermi level ($E_F$). Hole and electron dispersions are noticed around the Γ and M points, respectively. The results are consistent with the previous reports [22]. All the spectra were taken at 15 K in this work.

Figures 2(a) and 2(b) show the momentum-integrated TARPES intensity measured across the hole and electron FSs with pump fluences of 2.45 and 2.28 mJ/cm², respectively, as a function of pump-probe delays (Δt). Due to the poor efficiency of the HHG process, the energy resolution is set to be 250 meV. The integrated range is along the orange-solid and green-solid lines shown in Fig. 1(b) for the hole and electron FSs, respectively. After the intense pulse excitation, electrons are immediately excited above $E_F$ followed by relatively slow relaxation dynamics at both FSs. To see the photo-excited dynamics more clearly, we show in Figs. 2(c) and 2(d) integrated intensity above $E_F$ corresponding to the regions surrounded by green boxes in Figs. 2(a) and 2(b), respectively. Overall, immediate excitation and overshooting decay at $\Delta t = 0$ ps followed by relatively slow recovery dynamics are observed, reflecting the carrier dynamics. Additionally, oscillatory behaviors are clearly observed superimposed onto background carrier dynamics. These oscillatory components are especially evident at initial time ($\Delta t < 1.5$ ps). At later time of around $\Delta t = 3$ ps, on the other hand, it should be noticed that they exhibit a contrasting feature, *i.e.*, the photoemission intensity decreases at the hole FS, while the intensity at the electron FS increases. This behavior is also confirmed in time- and angle-resolved photoemission spectra shown in Fig. S2 as the downward band shift at $\Gamma$ point as well as the increase of electrons in electron pockets at M point. These signatures will be discussed later in more detail. To highlight oscillatory components, we first fit the background carrier dynamics with two-component exponential decay functions convoluted with a Gaussian. They are shown by black solid lines in Figs. 2(c) and 2(d), and then subtracted from experimental data. The oscillatory components are displayed in Figs. 2(e) and 2(f), respectively. These oscillations are in phase with each other, and cosine-like with the frequency of 5.3 THz. From the comparison with Raman spectra [23], this oscillation frequency is assigned to the $A_{1g}$ phonon mode, in which two Se layers oscillate symmetrically with respect to the sandwiched Fe layer as shown in Fig. 3(d). The recent results of Raman spectroscopy performed on single crystal FeSe revealed the frequency of the $A_{1g}$ phonon mode to be 5.5 THz, which justifies the assignment of the coherent phonon observed in the previous TARPES studies [19, 20]. Moreover, the fast Fourier transforms (FFTs) of the oscillatory components are shown in Fig. 2(g), and FFT peak amplitudes for both the hole and electron FSs significantly increases as a function of pump fluence shown in Fig. 2(h). These cosine-like and intensity-dependent behaviors confirm that the observed oscillations are attributed to the displacive excitation of coherent phonons (DECP) [24]. According to the DECP mechanism, photo-excitation moves the system to a free-energy curve of excited states with a minimum position at different $h$ from the equilibrium state. As a result, Se atoms simultaneously oscillate with a center at the new stable (*metastable*) position. Since the oscillatory components of the photoemission intensity above $E_F$ reflect the size of FSs [16], the cosine-like signature indicates the photo-

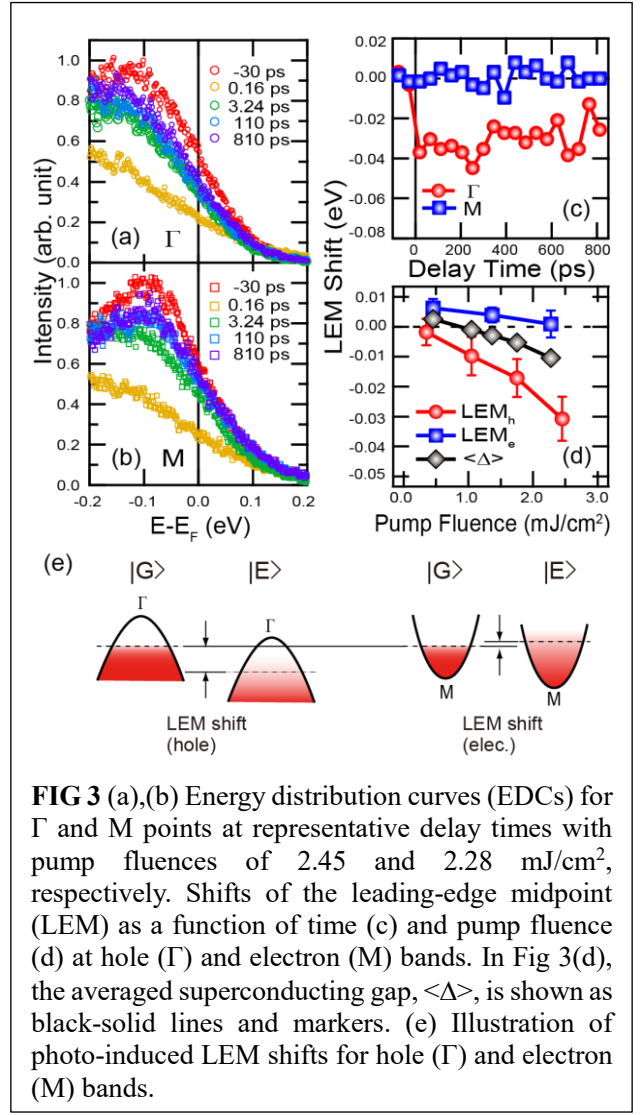

**FIG 3** (a),(b) Energy distribution curves (EDCs) for Γ and M points at representative delay times with pump fluences of 2.45 and 2.28 mJ/cm², respectively. Shifts of the leading-edge midpoint (LEM) as a function of time (c) and pump fluence (d) at hole (Γ) and electron (M) bands. In Fig 3(d), the averaged superconducting gap, <Δ>, is shown as black-solid lines and markers. (e) Illustration of photo-induced LEM shifts for hole (Γ) and electron (M) bands.

induced metastable state is toward smaller FSs for the hole and electron pockets.

**Long-lived charge disproportionate state**

As briefly seen, photoemission intensity exhibits contrasting features between the hole and electron FSs at larger delay time ($\Delta t = 3$ ps). This feature also persists at relatively long-delay time shown in Fig. S4. In order to investigate this behavior in more detail, we proceed to analyze energy distribution curves (EDCs) for slow dynamics. Figure 3(a) presents the EDCs before ($\Delta t = -30$ ps) and after ($\Delta t = 110$ and 810 ps) arrival of pump pulses for the hole FS, while the result for the electron FS is shown in Fig. 3(b). Pump fluences for the hole and electron bands are 2.45 and 2.28 mJ/cm², respectively. After photo-excitation, the EDCs for both the hole and electron FSs become significantly broader at $\Delta t = 0.16$ ps, yet return soon at $\Delta t = 3.24$ ps to almost the same as those at 110 and 810 ps. This indicates that the electronic temperature is well cooled. Regarding the shift of the EDCs, the clear one towards the lower energy side is noticed at the hole FS. At the electron FS

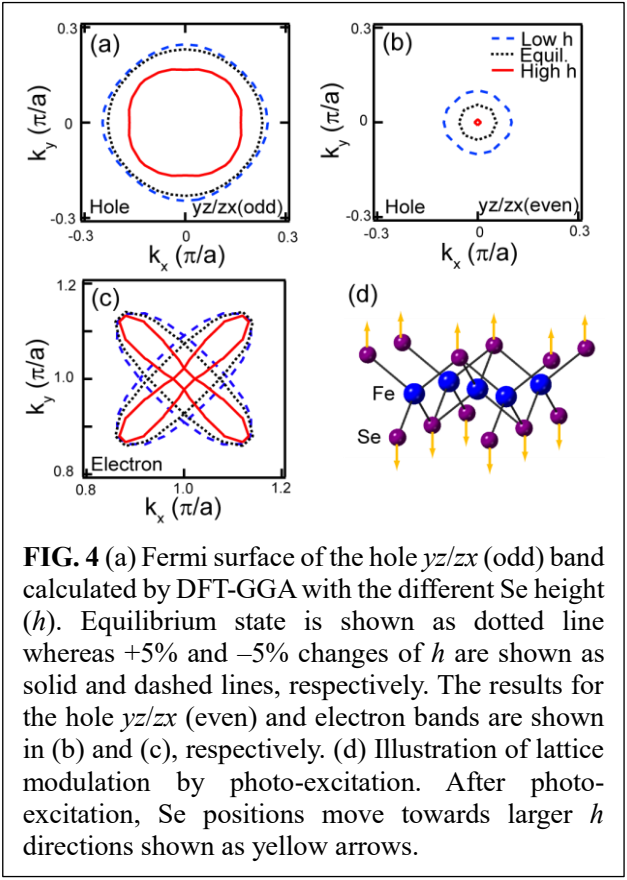

**FIG. 4** (a) Fermi surface of the hole *yz/zx* (odd) band calculated by DFT-GGA with the different Se height (*h*). Equilibrium state is shown as dotted line whereas +5% and –5% changes of *h* are shown as solid and dashed lines, respectively. The results for the hole *yz/zx* (even) and electron bands are shown in (b) and (c), respectively. (d) Illustration of lattice modulation by photo-excitation. After photo-excitation, Se positions move towards larger *h* directions shown as yellow arrows.

(Fig. 3(b)), on the other hand, the EDC intensities for $\Delta t$ = 110 ps and 810 ps at around $E - E_F$ = -0.1 eV are reduced while no clear shift appears. For quantitative insight, we evaluate the shift of the leading edge midpoint (LEM). Figure 3(c) shows the temporal LEM shifts. As is clearly seen, the LEM shift at the hole FS is negative while that at the electron FS is negligibly small. In other words, the temporal band filling is disproportionate between the hole and electron bands, as is illustrated in Fig. 3(e). Interestingly, the photo-excited electronic structure mimics that of the monolayer FeSe film, in which only electron FSs are observed [7]. Furthermore, this disproportionate band filling persists longer than the measured delay time of ~800 ps. Such a long lifetime of carriers can be ascribed to the indirect semimetallic band structures, where the electron-hole recombination must accompany with the assist of phonons with a large momentum [25, 26]. Figure 3(d) shows the LEM shift as a function of pump fluence, in which the values and estimated errors are the averages and standard deviations in $\Delta t$ >0, respectively. The trend of the disproportionality between the hole and electron bands becomes more evident with increasing pump fluence. However, it should be noticed that the negative shift of LEM for the hole FSs is not equal to the positive shift for the electron FSs, and this signature is more pronounced for higher fluence.

**Photo-induced superconductivity**

In a single hole and electron bands picture, photo-excited electrons are relaxed to the electron bands after the relatively fast processes of multiple electron-electron and electron-phonon scatterings. If the density of states (DOS) is similar between the hole and electron FSs, the LEM shift should be the same amount with the opposite sign. Hence, it should be unusual that the LEM shift at the electron FS is negatively small for higher fluence. If the overall shift of $E_F$ is included in these LEM shifts, one possibility is due to the surface photo-voltage (SPV) effect [27]. However, it is not expected to occur in a semimetallic system such as FeSe because the SPV effect is typically induced by the surface band bending of semiconductors. Another possibility is a multi-photon effect due to the strong excitation by near-infrared pump [28]. Since we confirmed the absence of photoelectron intensities due to a multi-photon effect by measuring no signal with only pump pulses, this explanation can also be unlikely. Floquet band theory may also explain our results, in which many replica bands appear apart from the original band by the photon energy in use for excitation. Although Floquet band theory significantly change the band structure, the reported Floquet states have relatively shorter lifetime around less than 1 ps [29-31]. Because our main focus in this work is LEM shifts at later than 100 ps, we have concluded that our results are less likely to be explained by Floquet theory. After considering all these effects, the overall LEM shifts should be ascribed to a gap originated from some orders.

In order to identify superconducting signatures more explicitly, we extract the averaged photo-induced superconducting gap, $\langle\Delta\rangle$, shown as a black-solid line in Fig. 3(d), which is given by the following relationship,

$$\langle\Delta\rangle = -\frac{m_h LEM_h + m_e LEM_e}{m_h + m_e}, \qquad (1)$$

where $m_h$ and $m_e$ are mass of the hole and electron pockets, respectively. Their ratio is $m_h/m_e = 2/3$ [32]. $LEM_h$ and $LEM_e$ are LEM shifts for the hole and electron pockets, respectively. Because $\langle\Delta\rangle$ negatively contributes to the LEM shifts, we plot -$\langle\Delta\rangle$ in Fig. 3(d). The detailed description on how to extract the averaged superconducting gap is found in Supplementary Information [33]. Considering that superconductivity coexists with the orbital ordering under equilibrium for FeSe, but the orbital ordering induces no band gap but a band splitting, photo-induced superconducting gap is the most plausible origin. The mechanism of the stabilization of the superconducting state due to the displacive excitation is explained in the next section.

**Lattice modulation induced by displacive excitations**

In order to determine whether *h* becomes higher or lower in the photo-induced metastable state [18], we performed band-structure calculations based on density functional theory (DFT). Results of the band-structure calculations are found in Supplemental Information [29]. Figures 4(a)-4(c) show the calculated FSs for the two hole bands ($yz/zx$(odd) and $yz/zx$(even)), and the electron band. Because DFT calculations for FeSe cannot provide quantitative agreement with the measured band dispersions [22, 34-35], band-dependent shifts of -0.08 and +0.17 eV, as well as renormalization with a factor of 3 and 2 are introduced for the electron and hole bands, respectively. Equilibrium state is shown as dotted lines whereas +5% and –5% changes of $h$ are shown as solid and dashed lines, respectively. Since the probe pulses are polarized along the detector slit in this work, the contribution to the photoemission intensity around the Γ point has been reported to be mainly from the $yz/zx$(even) orbtital due to the photoemission matrix element [22, 35]. In each band, the FS shrinks as $h$ increases. From the comparison between the experiments and calculations, higher $h$ is interpreted to be realized in the photo-induced metastable state, as illustrated as yellow arrows in Fig. 4(d). This trend agrees with the previous report measuring the band shift at the Γ point by high-energy-resolution TARPES [20], where initial dynamics of downward bands shift is revealed to be synchronized with the increase of $h$. Although the quantitative agreement is difficult to be achieved between DFT calculations and measured band dispersions, trend of the band shifts with respect to change of $h$ should be correct as reported previously [20], which directly measured the dynamics of band dispersions as well as lattice distortions, and compared them with DFT + DMFT calculations. Even though our method using DFT calculations is less quantitative for reproducing measured band dispersions than DFT + DMFT calculations, the fact that both methods predict the same tendency with respect to change of $h$ strongly suggests that our results of DFT calculations should be correct.

Regarding superconductivity, it has been reported that $T_c$ increases with higher $h$ by physical pressure [4]. Photo-excitation can induce the same tendency as the physical pressure and can be another tool for the enhancement of superconductivity. Successive recent reports on the photo-induced superconductivity in cuprate superconductors have discussed along the important role of lattice motions [11, 36]. For the excitation employed in most studies, mid-infrared pulse is used in order to resonantly excite the lattice degree of freedom (phonons). However, the eventual lattice modulation playing in a decisive role for a photo-induced superconductivity was found to be the Raman-active $A_g$ phonon, which is nonlinearly coupled to the photo-excited infrared-active phonon mode [11]. Thus, it is indicated that the important mechanism relies on how to access the $A_g$ lattice modulation that is favorable for the superconductivity. Interestingly, near-infrared pulses are also used for excitation, which initially excite the electronic degree of freedom. Although the precise mechanism is still under debate, it is proposed that the lattice modulation is induced by the change of the distribution of electronic system [37], which is very similar to our situation.

Along with cuprate superconductors, Fe-based superconductors can be an alternative candidate for photo-induced superconductivity because pronounced lattice motions are detected by coherent phonons, and interestingly the present results of higher $h$ in FeSe is favorable for the superconductivity. Long-lived aspect of photo-induced phase owing to indirect semimetalic band structures is also characteristic feature for Fe-based superconductors comparing with a short-lived feature for cuprate superconductors due to direct single-band structures.

In summary, we have investigated photo-excited electron dynamics of the hole and electron FSs in FeSe by TARPES using HHG technique. We have proposed that ultrafast optical technique can offer another route to create new long-lived electronic and lattice structures in FeSe as a consequence of its indirect semimetalic band structure.

**Method**
**Time- and angle-resolved photoemission spectroscopy**
TARPES is a pump-probe type measurement, where both pulses are originally generated from Ti:Sapphire amplification system with a repetition rate of 1 kHz and pulse duration of 35 fs. Near infrared (NIR) pulse (1.55 eV) is used for the pump while an extreme ultraviolet (XUV) pulses (27.9 eV) are used for the probe. XUV pulses are obtained by selecting the ninth order harmonic generated from Argon gas with using double-frequency pulse (3.10 eV). The time resolution is measured to be ~ 80 fs from the cross-correlation measurement between the pump and probe pulses, and the energy resolution is set to 250 meV. Photoemission spectra are measured using a Scienta R4000 hemispherical electron analyzer.

**Sample**
High quality single crystals of FeSe were grown by the chemical vapor transport method using $KCl/AlCl_3$ as transport agent [38]. Clean surfaces were obtained by cleaving *in situ*.

**Band structure calculations**
Band structure calculations based on density functional theory (DFT) were performed using a WIEN2k package [39]. $a$ = 3.7707 Å, $c$ = 5.521 Å, and $h$ = 1.4723 Å, where $h$ is the Se height from the Fe layer, are used for the lattice parameters in the equilibrium state [38].

**Author contributions**
T. Suzuki, T. Someya, T.H., S.M., M.W., K.O. performed TARPES measurements. T. Suzuki performed data analysis. M.F., T.K., N.I., and J.I.

conducted maintenance of HHG laser system and improvements of TARPES apparatus. S.K., Y.M., T. Shibauchi grew high-quality single crystals and characterized them. T. Suzuki, Y.M., T. Shibauchi, and K.O. wrote the manuscript. K.O. and S.S. designed the project. All the authors discussed the results and contributed to the manuscript.

**Data availability**

The data supporting the findings of this study are available from the corresponding author on request.

# Supplementary Information

## Band Dispersions for the Modulated Structures

Calculated band dispersions for the equilibrium structure and the modulated structure with ±5% variation of the Se height are shown in Fig. S1(a) and S1(b), respectively.

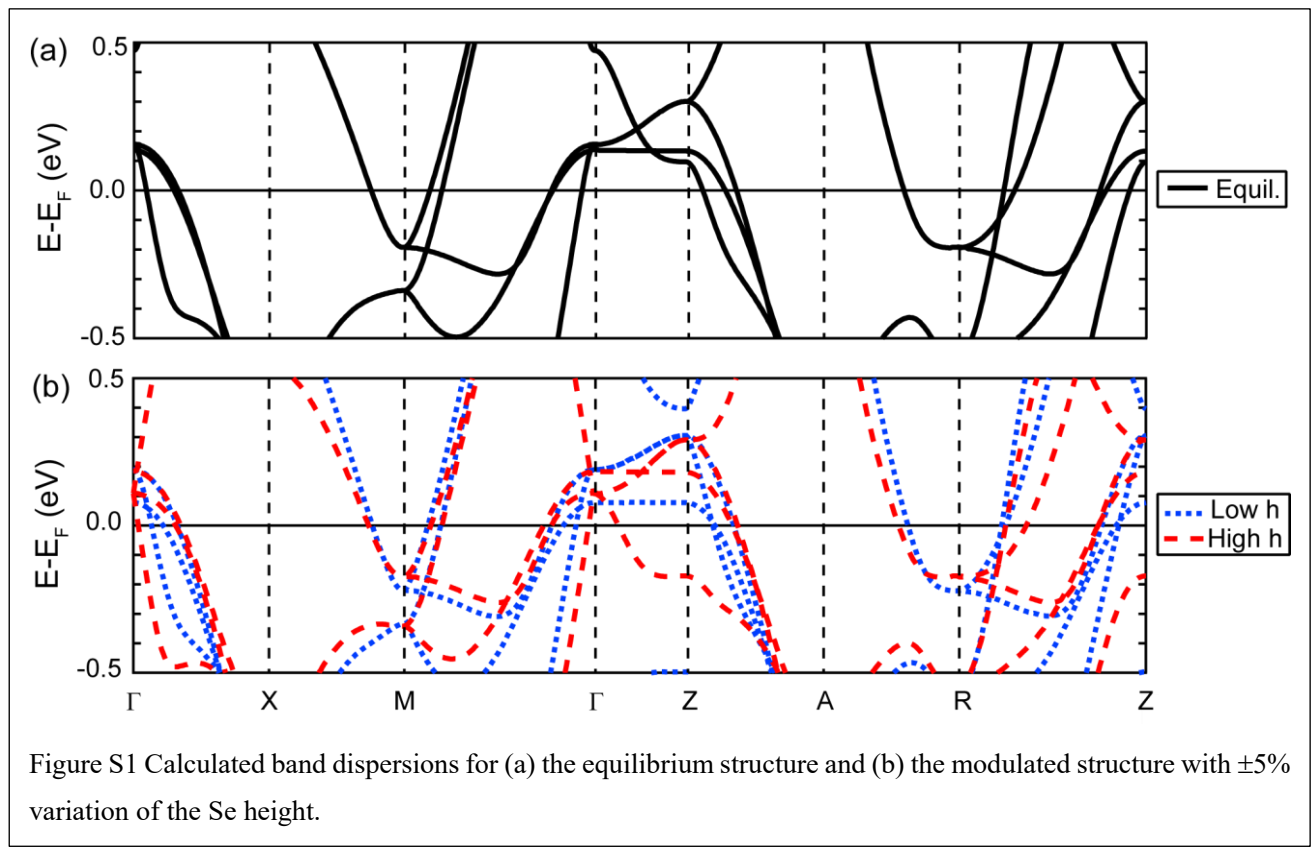

Figure S1 Calculated band dispersions for (a) the equilibrium structure and (b) the modulated structure with ±5% variation of the Se height.

# Time- and Angle-Resolved Photoemission Spectra

Time- and angle-resolved photoemission spectra at the Γ and M points are shown in Figs. S2(a)-S2(f) and Figs. S2(l)-S2(q), respectively. Difference spectra at the corresponding delay times are shown in Figs. S2(g)-S2(k) and Figs. S2(r)-S2(v) at the Γ and M points, respectively. They are obtained by subtracting Fig. S2(a) and S2(l), respectively. The black dashed line in Fig. S2(k) indicates the band downward shift at the Γ point while the black solid circles in Fig. S2 (v) show the increase of electrons in electron pockets at the M point.

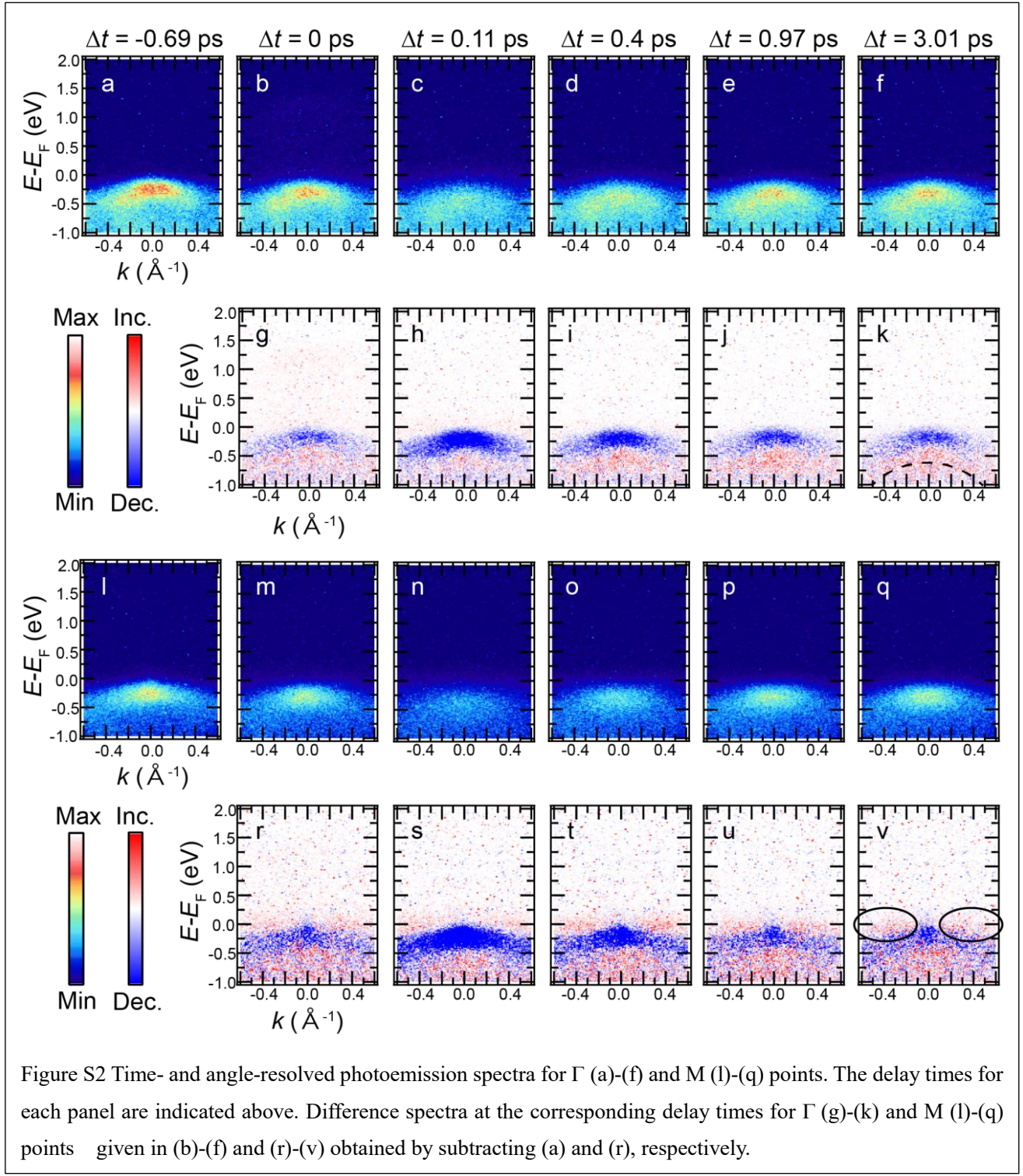

Figure S2 Time- and angle-resolved photoemission spectra for Γ (a)-(f) and M (l)-(q) points. The delay times for each panel are indicated above. Difference spectra at the corresponding delay times for Γ (g)-(k) and M (l)-(q) points given in (b)-(f) and (r)-(v) obtained by subtracting (a) and (r), respectively.

## Superconducting Gaps Extracted from LEM Shifts

Due to electron transfer from the hole to electron pockets, the decrease and increase of electron populations in the hole and electron pockets denoted as $\Delta N_h$ and $\Delta N_e$ respectively, are the same amount, and they are proportional to the mass of electrons. Given the chemical potential shifts denoted as $\Delta\mu_h$ and $\Delta\mu_e$, $\Delta N_h$ and $\Delta N_e$ are

$$|\Delta N_h| = |\Delta N_e|, \tag{1}$$

$$\Delta N_h = \left(\frac{1}{2\pi}\right)^2 \frac{2\pi m_h}{\hbar^2} \Delta\mu_h, \tag{2}$$

$$\Delta N_e = \left(\frac{1}{2\pi}\right)^2 \frac{2\pi m_e}{\hbar^2} \Delta\mu_e. \tag{3}$$

By considering superconducting gaps as well, additional shifts corresponding to the averaged superconducting gap $\langle\Delta\rangle$, LEM shifts for the hole and electron pockets, $LEM_h$ and $LEM_e$ are

$$LEM_h = \Delta\mu_h + \langle\Delta\rangle, \tag{4}$$

$$LEM_e = \Delta\mu_e + \langle\Delta\rangle. \tag{5}$$

From Eqs (1)-(5), $\langle\Delta\rangle$ is expressed by the LEM shifts as

$$\langle\Delta\rangle = \frac{m_h LEM_h + m_e LEM_e}{m_h + m_e}. \tag{6}$$

# TARPES Results for the Long-Delay Time Measurements

Time-resolved photoemission spectra for long-delay time measurements are shown in Figs. S4 (a) and S4 (b) taken at around the Γ and M points, respectively. Integrated photoemission intensities above $E_F$ corresponding to the surrounded region by green boxes in Figs. S4 (a) and S4 (b), respectively, are shown in Fig. S4 (c). Difference spectra are displayed in Figs. S4(d) and S4(e) for Γ and M points, respectively, obtained by subtracting spectra before arrival of pump. Integrated photoemission intensities above $E_F$ corresponding to the surrounded region by green boxes in Figs. S4 (d) and S4 (e), respectively, are shown in Fig. S4 (f).

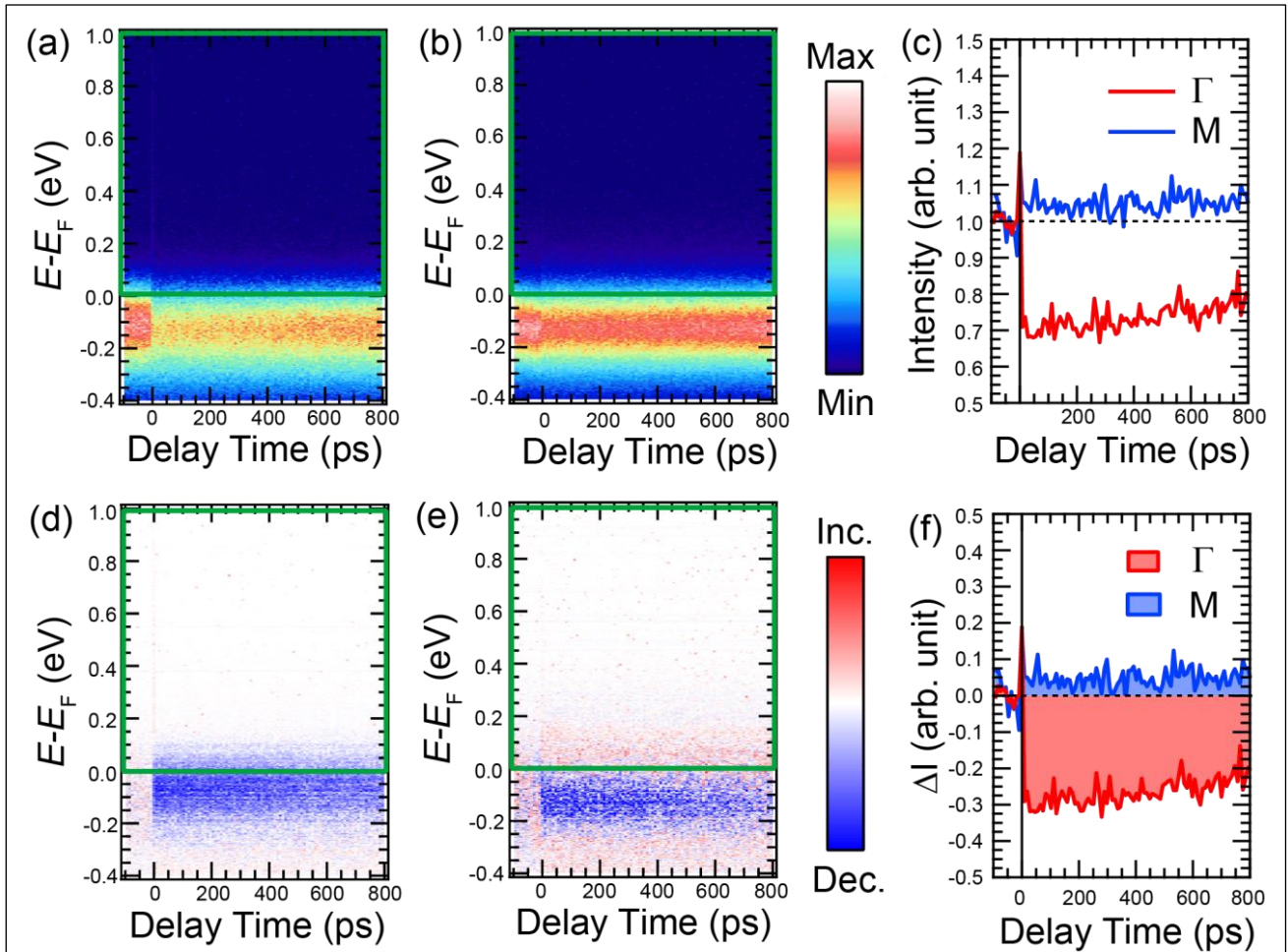

Figure S4  (a), (b) Time-resolved photoemission spectra for a long-delay time measurement taken at around the Γ and M points, respectively. (c) Integrated photoemission intensities above $E_F$ corresponding to the surrounded region by green boxes in (a) and (b), respectively. (c), (d) Difference spectra for Γ and M points, respectively, obtained by subtracting spectra before arrival of pump. (f) Integrated photoemission intensities above $E_F$ corresponding to the surrounded region by green boxes in Figs. S4 (d) and S4 (e), respectively.